\documentclass[aps,prl,preprint,groupedaddress]{revtex4}
\usepackage{graphicx}

\begin{document}

\title{Observation of a New Mechanism of Spontaneous Generation of Magnetic Flux in a
Superconductor}
\author{Ariel Maniv*, Emil Polturak, Gad Koren, Yuri Bliokh}
\affiliation{Physics Department, Technion - Israel Institute of Technology, Haifa
32000,Israel }
\author{Bj\"{o}rn Biehler, Brend-Uwe Runge, Paul Leiderer}
\affiliation{Physics Department, Konstanz University, Konstanz 75458,Germany }
\author{Boris Shapiro, and Irena Shapiro}
\affiliation{Physics Department, Bar-Ilan University, Ramat-Gan 52900,Israel }
\date{\today }

\begin{abstract}
We report the discovery of a new mechanism of spontaneous
generation of a magnetic flux in a superconductor cooled through
$T_c$. The sign of the spontaneous flux changes randomly from one
cooldown to the next, and follows a Gaussian distribution. The
width of the distribution increases with the size of the
temperature gradient in the sample. Our observations appear
inconsistent with the well known mechanisms of flux generation.
The dependence on the temperature gradient suggests that the flux
may be generated through an instability of the thermoelectric
superconducting-normal quasiparticle counterflow.
\end{abstract}


\maketitle

With the exception of ferromagnets, a spontaneous appearance of a
magnetic field in a physical system is a highly unexpected
phenomenon. Yet, such phenomenon was observed in superconductors
cooled through $T_c$
\cite{Tsuei_Kirtley,Kirtley5,AFM_Emil6,RazNature,Carmi_Emil_JJ,Monaco}.
In one case, the spontaneous magnetic field appeared as a
consequence of the d-wave symmetry of the order parameter of
HTSC\cite{Tsuei_Kirtley,RazNature}. In another case, the
spontaneous field was generated by cooling the superconductor
through $T_c$ under non equilibrium thermal
conditions\cite{AFM_Emil6,Carmi_Emil_JJ,Kirtley5, Monaco}. Here,
we report a new, unexpected appearance of a spontaneous field,
which occurs in a superconductor cooled in the presence of a
thermal gradient.

Although the effect described below appears completely unrelated,
the original motivation for this experiment followed our previous
work on the Kibble-Zurek cosmological
scenario\cite{Kibble1,Zurek2}. One of the key assumptions of this
scenario is that the temperature within the sample is uniform. The
limit on the size of $\nabla T$, the temperature gradient across
the sample, set by Kibble and Volovik\cite{Kibble_Volovik18}, is
that $\nabla T < T_{c}\hat{\varepsilon}/\hat{\xi}$. Here $T_{c}$
is the transition temperature,
$\hat{\varepsilon}=\frac{\hat{T}-T_c}{T_c}$ and $\hat{\xi}$ are
the reduced temperature and coherence length respectively at the
temperature $\hat{T}$ at which fluctuations of the order parameter
return to thermal equilibrium\cite{Zurek2}. Our experiment was
designed to see what happens to the formation of topological
defects once this criterion is not satisfied.

The experimental setup is the same as described in Ref. 4, with
the exception of a non-uniform heating, generating intentional
temperature gradients in the sample. Briefly, our samples were 300
nm thick c-axis oriented YBa$_{2}$Cu$_{3}$O$_{7}$ films with
$T_{c}$ $\simeq$ 90 K, grown on a SrTiO$_{3}$ substrate. The
samples were placed atop the sensing coil of a HTSC SQUID
magnetometer. In our arrangement the SQUID remains at a
temperature of 77 K, and is not affected by the temperature of the
sample which can be heated and cooled independently. The film is
heated above $T_{c}$ using a light source and cools by exchanging
heat with its environment. The light source is a pulsed YAG
laser\cite{Leiderer}. Single pulses (FWHM $\sim 10$ ns) were used
to heat the film. The laser pulse passes through the substrate and
illuminates \emph{non-uniformly} a selected area of the film. At a
laser wavelength of 1.06 $\mu $m, the SrTiO$_{3}$ substrate is
transparent and practically all the light is absorbed in the film.
Hence, only the film heats up, while the substrate remains near
the base temperature of 77 K. The 1 mm thick substrate has a heat
capacity about 10$^{3}$ larger than that of the film. The heat
from the film escapes into the substrate, which acts as a heat
sink. This small thermal mass of the film allows us to achieve
cooling rates in excess of 10$^{8}$ K/sec. The cooling rate at
$T_{c}$ can be varied by changing the amount of energy delivered
by the laser pulse. The system is carefully shielded from the
earth's magnetic field, with a residual field of less than 50
$\mu$G. Additional small coil adjacent to the sample was used to
test the field dependence of the results, at fields ranging from
less than 50 $\mu$G up to 60 mG.

Non-uniform illumination was generated either by using a
non-uniform light beam, or by covering some part of the sample. An
example of one such arrangement is shown in the inset of figure
\ref{fig:Distribution}. Here, the strongly illuminated area is a
stripe across the film. In another configuration, the perimeter of
the film was masked, while an area of 4mm in diameter in the
center was exposed to the beam. Qualitatively, the results
presented here do not depend on the exact illumination profile.
Under such non-uniform illumination, the film cools down in a two
stage process. In the first stage, the heat deposited by the laser
pulse in the film is dumped into the SrTiO$_{3}$ substrate on a
time scale of a $\mu$s. As a result, the temperature of the part
of the substrate closest to the illuminated area increases by up
to 5K above 77K, depending on the laser energy. In the second
stage, heat is transferred from the hotter part of the substrate
to its cold parts on a time scale of several tens of ms. During
all this time temperature gradients are present across the sample.
The relatively slow time scale on which the substrate cools is due
to its thermal mass, which is much larger than that of the film.

Previously, under homogeneous illumination ($\nabla T \sim$
1K/cm), we have observed the generation of spontaneous flux during
a rapid quench of a superconducting film\cite{AFM_Emil6}. The flux
appeared faster than the temporal resolution of our SQUID, which
is $\sim 10 \mu$s. In the following, we refer to this signal as
the "fast" signal. The polarity of the flux from one quench to the
next was random, following approximately a Gaussian distribution
centered at zero. The width of this distribution increased weakly
with the quench rate, a result which is broadly consistent with
the Kibble-Zurek scenario.

Under non-homogeneous illumination, we estimate that $\nabla T$
increased to about 300 K/cm for the largest pulse energy used.
This is still less than the limit set by the homogeneous
criterion\cite{Kibble_Volovik18} of 10$^{4}$ K/cm. Under these
conditions, the "fast" signal showed no appreciable change.
However, in addition to the "fast" signal, an unexpected, much
larger signal has appeared after a relatively long delay of 1-10
ms (see figure \ref{fig:Slow_Fast}). This signal was completely
absent during measurements using a homogeneous illumination. We
point out that the time it takes to cool below $T_{c}$ is on the
order of 1 $\mu$s. Consequently, the "slow" signal appears while
the film is already in the superconducting state.

The polarity of the non-homogeneous, "slow" signal was also random
from one quench to the next. Similarly to the "fast" signal, the
amount of flux generated in a given quench followed a Gaussian
distribution centered at zero. This is shown in Figure 2. However,
the amount of flux associated with the "slow" signal is larger
than that of the "fast" signal by an order of magnitude (see also figure \ref%
{fig:Signal Vs._Energy per pulse}).

After analyzing data aquired using different pulse energies, we found
that the amount of spontaneous flux, characterized by the distribution
width, increases with the pulse energy. This contrasts the results found
under the conditions of uniform illumination (there the distribution width
decreased with increasing pulse energy). This is clearly seen in figure \ref%
{fig:Signal Vs._Energy per pulse}, in which the signal dependence on pulse
energy is shown. Note that increasing the pulse energy also increases the
thermal gradients generated across the film.

Finally, measurements were repeated under different external
magnetic fields ranging from less than 50 $\mu$G up to 60 mG. As
figure \ref{fig:Signal Vs._Energy per pulse} clearly shows, the
results do not depend on the external field.

The results at non-homogeneous conditions point towards two
important conclusions. First, as already noted above, increasing
the temperature gradients across the film by 2 orders of magnitude
(from 1 K/cm up to 300 K/cm) does not change the "fast" signal.
Therefore we conclude that the homogeneous
approximation\cite{Kibble_Volovik18} indeed holds, at least for
thermal gradients up to $\sim 10^{2}$ K/cm. Second, the dependence
of the "slow" signal on pulse energy and the long time scale
clearly imply that it originates from another mechanism, rather
than the Kibble-Zurek scenario. In the following, we discuss
several other mechanisms which may generate magnetic flux, and
examine their possible relevance to our observations.

The Hindmarsh-Rajantie model\cite{HR_model} predicts a conversion
of thermal energy into magnetic field fluctuations while the
sample is in the critical region near $T_c$. In our experiment,
the sample passes through this region in less than 1 $\mu$s, while
the slow signal develops on a time scale 3 to 4 orders of
magnitude slower, 1 - 10 ms. So, this scenario does not fit with
our observations.

Another possibility is a change in the spatial distribution of
residual magnetic flux inside the film. Re-arrangement of magnetic
flux lines can happen during partial illumination of the samples.
Magnetic flux can move in or out of the heated part of the film,
changing the magnetic flux distribution. Re-distribution of
magnetic flux can then change the actual amount of flux coupled to
the SQUID, even though the net change is zero. We investigated
this mechanism in separate measurements done at the university of
Konstanz, Germany using a magneto-optic system capable of sub ns
resolution \cite{Bolz}. We found that re-distribution of flux
takes place within several ns, which again is inconsistent with
the time scale of our slow signal. In addition, re-distribution of
flux should depend on the ambient field, which is not borne by our
data.

Several theory papers\cite{Kopnin,Shapiro} proposed that flux can
be generated by an instability of a propagating
normal-superconducting phase boundary front, which indeed is
present in our samples as the film cools after a non-homogeneous
heating pulse. If this mechanism is viable, it should act during
less than 1 $\mu$s after the heating pulse, since at later times
the entire film cools back into the superconducting state and the
front disappears. Again, this is 3 order of magnitude faster than
the time after which the slow signal is observed.

Spontaneous flux can be formed at large angle grain
boundaries\cite{Tsuei_Kirtley,Mannhart} as a consequence of the
d-wave symmetry of the order parameter.  Our samples are epitaxial
thin films, in which large angle grain boundaries are absent.
Hence, this mechanism can not explain the origin of our signal.

One clue as to the origin of the effect comes from the observation that the
temperature gradients across the sample relax on the same time scale as the
time over which the slow signal develops. Therefore it is natural to
associate it with some thermoelectric effect. This association would also be
consistent with the size of the effect increasing with the energy deposited
in the film. Thermo-electric effects (the Seebeck effect or the Nernst effect%
\cite{Poole}) can generate flux lines as a result of
superconducting currents in the film.

In superconductors, the Nernst effect is a result of the motion of flux
lines along the thermal gradient. Clearly, this effect depends on the
ambient magnetic field. Since we see no such dependence, we conclude that
the Nernst effect does not explain our measurements.

Regarding the Seebeck effect in a superconductor, thermal gradients produce
a counterflow of normal quasiparticles and Cooper pairs. The net electric
current is zero\cite{VanHarlingen}. However, as noted by Ginzburg\cite%
{Ginzburg}, in some cases such thermo-electric currents can
generate magnetic flux. One example is the anisotropic
thermo-electric effect\cite{Ginzburg}, in which the supercurrent
and the normal current are not co-linear and form a current loop.
This happens if the Seebeck coefficient is anisotropic and the
direction of the thermal gradient is not parallel to one of the
superconductor's symmetry axes. Then, the superconducting
countercurrent does not exactly cancel the normal current at every
point of the film, hence generating a non-zero magnetic flux.
Measurements done by Subramaniam et al.\cite{Subramaniam} show
that for \textit{untwinned} YBCO crystals, thermoelectric
properties are indeed anisotropic. However, our films are twinned,
so there is no anisotropy between the \textbf{a} and \textbf{b}
directions (parallel to the surface of the film). Under a
temperature gradient of 300 K/cm, we estimate the thermo-electric
current $\mathit{I}\sim 5\times10^{-5}$A. This estimate is based
on the measured thermal coefficients\cite{Subramaniam}.

If the spontaneous flux was generated via a linear thermoelectric
effect, we would expect the polarity of the flux generated to be
the same in each measurement, since the temperature gradient in
the sample is nominally the same. Since the polarity of the
measured flux is random in each measurement, this suggests that
perhaps an instability occurs.

One well known example is the plasma ''two stream
instability''\cite{Plasma}. In a plasma, the Lorentz force between
the two opposing electron beams vanishes only if the currents
cancel exactly everywhere. With spatial current fluctuations, the
cancellation does not hold, resulting in a net repulsive force
between the currents. This in turn leads to further separation of
the currents, creating a current loop and a magnetic flux. This
situation is illustrated in figure \ref{fig:Counterflow
Instability}. The sense of the current in the loop (clockwise or
counterclockwise) is random, having been determined by the initial
fluctuation which separates the current and countercurrent. This
is in line with our data. In superconductors, the counterflow
consists of a normal current opposed by a supercurrent. A high
frequency plasma instability in a superconductor was predicteded
by Kempa et al. \cite{SC_Plasma}. Another variant of the ''two
stream instability'' was proposed by Bliokh and
Shapiro\cite{Plasma_Private_Comm}, who showed that in the
framework of the two-fluid model\cite{Tinkham100}, a uniform
superconducting-normal quasiparticle countercurrent is unstable
with respect to spatial fluctuations. The analysis reveals that
fluctuations in the current density can induce a \emph{low
frequency} instability and generate a magnetic field. Denoting the
velocities and densities of the normal and superconducting
components by $V_{n,s}$ , $n_{n,s}$, and taking
$n_{s}V_{s}=n_{n}V_{n}$ and $n_{s}+n_{n}=n$, the growth rate of
the unstable mode has the form\cite{Plasma_Private_Comm}:

\begin{equation}
\omega =\frac{n^{2}}{n_{n}n_{s}}\frac{V_{s}^{2}k^{2}}{\nu }
\end{equation}

where $\nu $ is the electron relaxation time and $k$ is the
wavevector of the unstable mode. To obtain a numerical estimate,
we determine $V_{s}$ from $J_s= en_s V_s$, using the
thermoelectric current estimated above, $\mathit{I}\sim
5\times10^{-5}$A and the sample cross section. The wavevector $k$
is determined by taking the size of a fluctuation as $\lambda$,
the penetration depth. This choice fixes $k=2\pi /\lambda$. For
other quantities in (1), we used the two fluid model expressions,
a charge density $n$ of 10$^{21}$ holes/cm$^3$, and $\nu\sim
10^{14}$ Hz. Assuming the current flows at the final stage of the
experiment near the boundary of the sample, we find $\omega \sim$
10$^{+3}$Hz- 10$^{+4}$Hz, which is consistent with the measured
experimental growth rate (ranging between 10 Hz - 10$^{3}$Hz).
Hence, this scenario gives a possible explanation for the origin
of the measured spontaneous flux.

In conclusion, we have discovered a new mechanism of spontaneous
flux generation in a superconductor quenched through $T_c$ in the
presence of a temperature gradient. One mechanism which may be
responsible for this new effect is an instability of the
thermoelectric current distribution.

\begin{acknowledgments}
We thank M. Ayalon, L. Iomin, and S. Hoida for technical
assistance. This work was supported in part by the Israel Science
Foundation (Grant No. 1565/04), by the Heinrich Hertz Minerva
Center for HTSC, and by the Fund for Promotion of Research at the
Technion. The work in Konstanz and Bar-Ilan was supported by the
German-Israeli Foundation.
\end{acknowledgments}

*corresponding author: mariel@tx.technion.ac.il

\begin{figure}[h!]
%
%\includegraphics[scale=0.4,angle=-90]{Figure1.pdf}
%
%\framebox[3in]{\rule[1.125in]{0in}{1.125in}}
%\makebox[5in]{\rule[1.125in]{0in}{1.125in}}
\caption{Typical SQUID signals showing the fast (top trace) and
slow(bottom trace) formation of spontaneous flux (note the
different scale of the horizontal axes.) Arrows show the time at
which the laser pulse was applied.} \label{fig:Slow_Fast}
\end{figure}

\begin{figure}[h!]
%
%\includegraphics[width=2.5in,height=2.5in]{Figure2.pdf}
%
%\framebox[3in]{\rule[1.125in]{0in}{1.125in}}
%\makebox[5in]{\rule[1.125in]{0in}{1.125in}}
\caption{Typical distribution of spontaneous flux under
non-homogeneous illumination. Solid black bars show the noise
distribution, while the dashed curve shows a Gaussian fit to the
signal distribution. The inset shows typical non-homogeneous
illumination profile.} \label{fig:Distribution}
\end{figure}

\begin{figure}[h!]
%
%\includegraphics[scale=0.3,angle=-90]{Figure3.pdf}
%\includegraphics[height=2in][width=3in]{Figure3.pdf}}
%
%\framebox[3in]{\rule[1.125in]{0in}{1.125in}}
%\makebox[5in]{\rule[1.125in]{0in}{1.125in}}
\caption{Signal distribution as a function of pulse energy,
showing the difference between homogeneous and non-homogeneous
illumination. Also shown are measurements done at several
different external magnetic fields. The error bars are
statistical.} \label{fig:Signal Vs._Energy per pulse}
\end{figure}

\begin{figure}[h!]
%
%\includegraphics[scale=0.3,angle=-90]{Figure4.pdf}
%
%\framebox[3in]{\rule[1.125in]{0in}{1.125in}}
%\makebox[5in]{\rule[1.125in]{0in}{1.125in}}
\caption{A schematic picture of the current loop formed by the
super and normal thermoelectric currents, separated as a result of
the instability. } \label{fig:Counterflow Instability}
\end{figure}

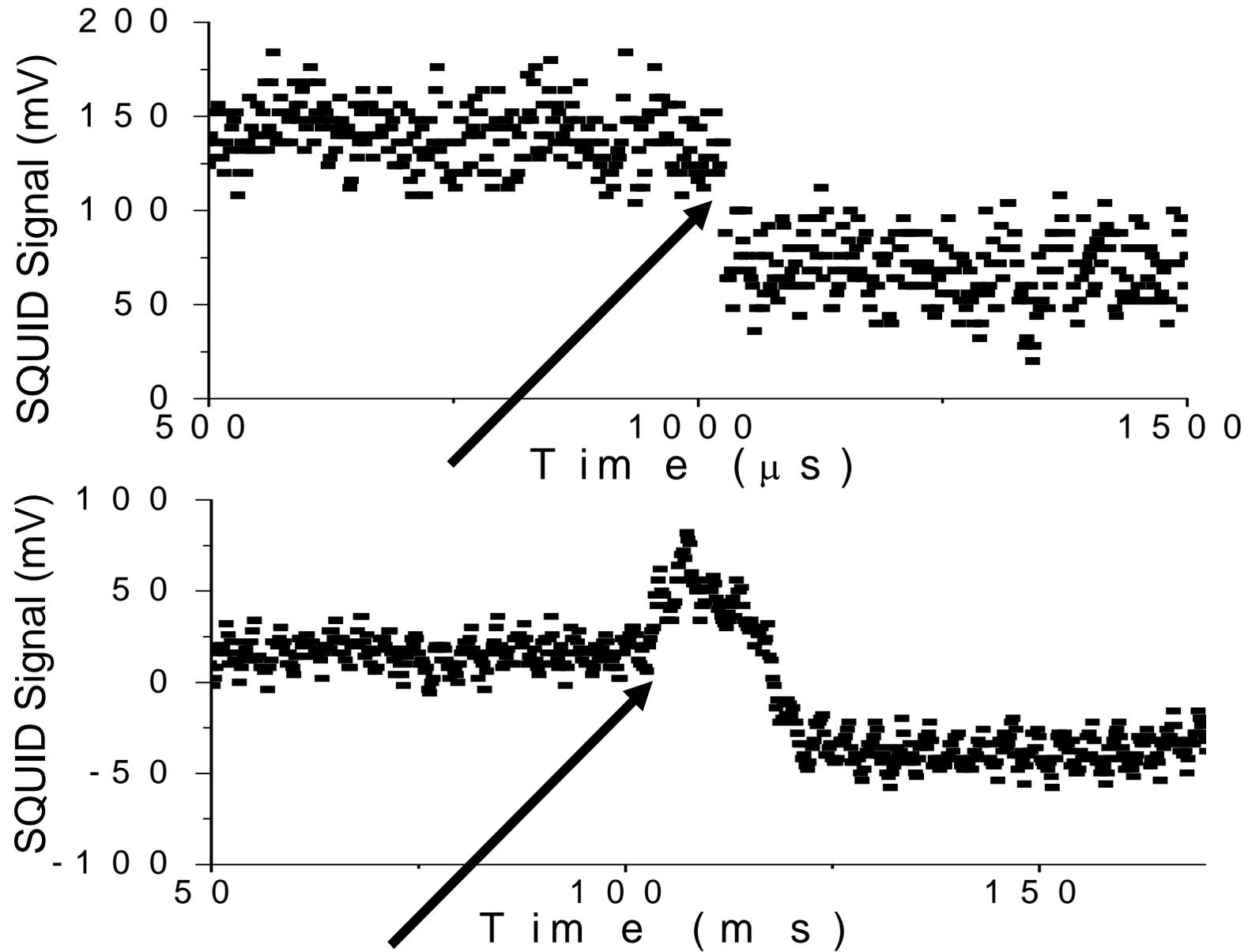
\end{document}